\documentclass[aip, jcp, reprint, twocolumn]{revtex4-1}
\usepackage[T1]{fontenc} 
\usepackage{amsmath}
\usepackage{amsfonts}
\usepackage{ucs}
\usepackage{graphicx,bm,}
\usepackage[caption=false]{subfig}
\usepackage{breqn}

\makeatletter
\let\cat@comma@active\@empty
\makeatother

\newcommand{\angstrom}{\textup{\AA}}
\newcommand{\eps}{\epsilon}

\begin{document}
\title{Wave function methods for canonical ensemble thermal averages in correlated many-fermion systems}
\author{Gaurav Harsha}
\email{gauravharsha05@gmail.com}
\affiliation{Department of Physics and Astronomy, Rice University, Houston TX 77005}
\author{Thomas M. Henderson}
\affiliation{Department of Physics and Astronomy, Rice University, Houston TX 77005}
\affiliation{Department of Chemistry, Rice University, Houston TX 77005}
\author{Gustavo E. Scuseria}
\affiliation{Department of Physics and Astronomy, Rice University, Houston TX 77005}
\affiliation{Department of Chemistry, Rice University, Houston TX 77005}

\begin{abstract}
  We present a wave function representation for the canonical ensemble thermal density matrix by projecting the thermofield double state against the desired number of particles. The resulting canonical thermal state obeys an imaginary time-evolution equation. Starting with the mean-field approximation, where the canonical thermal state becomes an antisymmetrized geminal power wave function, we explore two different schemes to add correlation: by number-projecting a correlated grand-canonical thermal state, and by adding correlation to the number-projected mean-field state. As benchmark examples, we use number-projected configuration interaction and an AGP-based perturbation theory to study the Hydrogen molecule in a minimal basis and the six-site Hubbard model.
\end{abstract}

\maketitle

\section{Introduction}

Thermal properties of many-body systems can be computed either in the canonical ensemble or the grand-canonical ensemble. The choice of ensemble makes no practical difference in the final result in large systems. It does so, however, for a finite system. This is because the relative fluctuation in particle number in the grand-canonical ensemble scales as the inverse square root of particle number itself, i.e.
\begin{equation}
  \frac{
    \sqrt{\langle N^2 \rangle_{gc} - \langle N \rangle_{gc}^2}
  }{
    \langle N \rangle_{gc}
  } \sim \frac{1}{\sqrt{\langle N \rangle_{gc}}},
\end{equation}
and vanishes in the limit $\langle N \rangle_{gc} \rightarrow \infty$, where $\langle \ldots \rangle_{gc}$ denotes the grand-canonical thermal expectation value.

A wide range of methods are available to study the thermal properties of quantum systems within the grand-canonical ensemble, e.g., thermal Hartree-Fock,~\cite{mermin_stability_1963, sokoloff_consequences_1967} perturbation theories,~\cite{matsubara_new_1955, santra_finite-temperature_2017, hirata_converging_2018} path integral and Green's function methods,~\cite{zgid_finite_2017} finite-temperature Quantum Monte Carlo (QMC),~\cite{zhang_finite_temperature_1999, militzer_path_2000, rubenstein_finite-temperature_2012, schoof_ab_2015, takai_finite-temperature_2016, claes_finite-temperature_2017, liu_ab_2018, he_finite-temperature_2019, petras_using_2020, liu_unveiling_2020} density matrix renormalization group and density functional theory based methods,~\cite{verstraete_matrix_2004, feiguin_finite-temperature_2005, stoudenmire_minimally_2010, pittalis_exact_2011, nocera_symmetry-conserving_2016, ren_time-dependent_2018} as well as the more recently explored thermal equivalents of configuration interaction and coupled cluster,~\cite{sanyal_thermal_1992, sanyal_systematic_1993, mandal_thermal_1998, mandal_finite-temperature_2003, hermes_finite-temperature_2015, hummel_finite_2018, white_time-dependent_2018, harsha_thermofield_2019, harsha_thermal_cc_2019, shushkov_real-time_2019, white_time-dependent_2019, white_finite-temperature_2020} and algorithms for quantum computers.~\cite{wu_variational_2018, mcardle_variational_2019, zhu_variational_2019, motta_determining_2020}

In contrast, canonical ensemble techniques are scarce and even fewer are suitable for efficient application to correlated electronic systems. One way to enforce a fixed number of particles is by introducing a second Lagrange multiplier $\mu_2$ for the fluctuation, in much the same spirit as the chemical potential $\mu_1$ acts as a Lagrange multiplier to fix the number of particles. That is, one can either define a generalization of the density operator as
\begin{equation}
  \rho = \exp \left[ -\beta \left( H - \mu_1 (N - N_0) - \mu_2 (N^2 - N_0^2) \right) \right ],
\end{equation}
where the parameters $\mu_1$ and $\mu_2$ enforce the constraints,
\begin{equation}
  \langle N \rangle = N_{0},
  \quad \mathrm{and,} \quad
  \langle N^2 \rangle = N_0^2,
\end{equation}
or introduce corrections to the grand-canonical ensemble averages by subtracting contributions from wrong number sectors in the Hilbert space.~\cite{kosov_calculations_2008} While this provides the convenience of using several available grand-canonical methods, such simultaneous optimization problems can be numerically tedious as the optimized values of $\mu_2$ are generally very large and ideally infinite, something which has also been observed in spin-projection.~\cite{andrews_spin_1991} On the other hand, we can evaluate the ensemble averages in the appropriate number sector to begin with, e.g. in the minimally entangled typical thermal states algorithm,~\cite{stoudenmire_minimally_2010, binder_symmetric_2017} canonical ensemble perturbation theory,~\cite{jha_finite-temperature_2020} and projection based techniques.\cite{tanabe_quantum_2005, esashika_effects_2005, nakada_new_2006, magnus_quantum_2017}

For a wide variety of problems which involve isolated finite systems with a fixed number of particles, the canonical ensemble is more appropriate. Examples of such systems include molecules in a warm gaseous phase (of interest in geochemistry),~\cite{guillot_interiors_1999} ultra-cold chemical systems,~\cite{balakrishnan_perspective:_2016, bohn_cold_2017} quantum wires with number conserving Majorana modes,~\cite{diehl_topology_2011, ortiz_many-body_2014, iemini_localized_2015} and superconductivity in small grain systems.\cite{mastellone_small_1998} Besides, the canonical ensemble provides a potential computational advantage over grand canonical alternatives since it eliminates the need for finding the appropriate chemical potential. Evidently, a robust and convenient framework to study canonical-ensemble finite-temperature properties of finite many-body fermionic systems is desirable.

In this manuscript, we leverage the thermofield dynamics\cite{matsumoto_thermo_1983, semenoff_functional_1983, umezawa_methods_1984, evans_heisenberg_1992} to construct a number-projected thermal wave function, called the canonical thermal state, which provides an exact wave function representation of the canonical ensemble density matrix. It obeys an imaginary-time Schr\"odinger equation which can be solved at various levels of approximation, and at the level of mean-field, reduces to a number-projected BCS wave function, also known as the antisymmetrized geminal power (AGP) state.\cite{coleman_structure_1965} A similar number-projected BCS theory for the canonical thermal state was also proposed by the authors of Refs.~\onlinecite{tanabe_quantum_2005, esashika_effects_2005, nakada_new_2006}. Mean-field description, however, misses out on a lot of important physics. Here, we provide a recipe to generalize correlated ground-state theories (e.g., perturbation theory, CI, CC, etc.) to finite-temperature. Moreover, the identification of the mean-field state as an AGP allows us to exploit the newly developed tools for efficient evaluation of the thermal expectation values via AGP density matrices.~\cite{khamoshi_efficient_2019} We restrict our discussion to electronic systems, but generalization to other fermionic and bosonic systems is straightforward.

\section{Thermofield dynamics}

Thermofield dynamics is conventionally formulated for the grand-canonical ensemble, where it constructs a wave function representation of the thermal density operator by introducing a conjugate copy of the original system such that the ensemble thermal averages can be expressed as an expectation value over the thermal state,
\begin{equation}
  \langle \mathcal{O} \rangle = \mathrm{Tr}\left( e^{-\beta(H - \mu N)} \mathcal{O} \right)
  = \frac{\langle \Psi(\beta) \vert \mathcal{O} \vert \Psi (\beta) \rangle}{\langle \Psi (\beta) \vert \Psi (\beta) \rangle},
\end{equation}
where the thermal state $\vert \Psi (\beta) \rangle$ is given by
\begin{subequations}
  \begin{align}
    \vert \Psi (\beta) \rangle &= e^{-\beta (H - \mu N) / 2} \vert \mathbb{I} \rangle,
    \\
    \vert \Psi (0) \rangle &= \vert \mathbb{I} \rangle = \prod_{p} \Big ( 1 + c_p^\dagger \tilde{c}_p^\dagger \Big ) \vert -; - \rangle.
  \label{inf-temp-gcan-thermal-state}
  \end{align}
\end{subequations}
Here $\beta$, $\mu$, $H$ and $N$ are the inverse temperature, chemical potential, the Hamiltonian, and the number operator respectively. The identity state $\vert \mathbb{I} \rangle$ is the exact infinite-temperature thermal state and is an extreme BCS state with Cooper pairs formed by pairing physical particles with the corresponding conjugate particles. The norm of the state gives the partition function.
The product in Eq.~\ref{inf-temp-gcan-thermal-state} runs over all spin-orbitals $p$ and $\vert -; -\rangle$ denotes the vacuum state for both the physical and conjugate systems.
By its definition, the thermal state obeys imaginary-time evolution equations, one each for $\beta$ and $\mu$,
\begin{subequations}
  \begin{align}
    \frac{\partial}{\partial \beta} \vert \Psi (\beta) \rangle &= -\frac{1}{2} H \vert \Psi(\beta) \rangle,
    \\
    \frac{\partial}{\partial \mu} \vert \Psi (\beta) \rangle &= \frac{\beta}{2} N \vert \Psi(\beta) \rangle,
  \end{align}
\end{subequations}
where we have assumed that $[H, N] = 0$, as \textit{ab-initio} electronic systems are number-conserving.

Like the ground state, finding $\vert \Psi (\beta) \rangle$ exactly is possible only for very small systems with a few electrons, and suitable approximations are generally required. The simplest approximation is the mean-field approach, where $H$ is replaced with a one-body mean-field Hamiltonian $H_0$. In the basis where $H_0 = \sum_p \eps_p c^\dagger_p c_p$, the resulting mean-field thermal-state is a BCS state of the form
\begin{align}
  \vert 0 (\beta, \mu) \rangle &= e^{-\beta (H_0 - \mu N) / 2} \vert \mathbb{I} \rangle,
  \nonumber
  \\
  &= \prod_p \Big( 1 + e^{-\beta (\eps_p - \mu) / 2} c^\dagger_p \tilde{c}^\dagger_p\Big) \vert -;- \rangle.
\end{align}
Higher order approximations are generally formulated with the mean-field state as the reference,
\begin{equation}
  \vert \Psi (\beta) \rangle \simeq \Omega(\beta, \mu) \, \vert 0 (\beta, \mu) \rangle,
\end{equation}
which resembles the interaction picture approach. We exploited this theory in Refs.~\onlinecite{harsha_thermofield_2019, harsha_thermal_cc_2019} to formulate finite-temperature versions of configuration interaction and coupled cluster theory. We recommend these articles and references therein for further details on thermofield theory.

\section{Canonical ensemble theory}

The canonical ensemble thermal state can be constructed by projecting the grand-canonical state against the desired particle number $N_0$,
\begin{equation}
  \vert \Psi (\beta) \rangle_{c}
  = \mathcal{P}_{N_0} \vert \Psi (\beta) \rangle_{gc},
\end{equation}
where $\mathcal{P}_{N_0}$ projects $\vert \Psi(\beta) \rangle_{gc}$ onto the Fock-space with $N_0$ electrons.
The particle-conserving property of $H$ implies that $[H, \mathcal{P}_{N_0}] = 0$, and the resulting canonical thermal state obeys an imaginary-time evolution equation analogous to its grand-canonical counterpart,
\begin{equation}
  \frac{d}{d \beta} \vert \Psi (\beta) \rangle_{c} = -\frac{1}{2} H \vert \Psi(\beta) \rangle_{c}.
  \label{imag-time-evol-eq}
\end{equation}
Like the grand-canonical theory, a series of approximations can be introduced, from a simple mean-field theory to higher order theories that add correlation effects on it.

\subsection{Mean-field formalism}
The imaginary-time evolution equation can be integrated within the mean-field approximation, $H \approx H_0$. As for the grand-canonical theory, using an $H_0$ that carries no implicit temperature dependence, and working in a basis where it is diagonal, the mean-field state becomes
\begin{subequations}
  \label{mean-field-eqns}
  \begin{align}
    \vert \Psi_{0} (\beta) \rangle_{c}
    &= \mathcal{P}_{N_0} \vert 0(\beta, \mu=0) \rangle,
    \label{mean-field:eqn-a}
    \\
    &=
    \mathcal{P}_{N_0} \prod_{p} \Big ( 1 + e^{-\beta \eps_p / 2} c_p^\dagger \tilde{c}_p^\dagger \Big ) \vert -; - \rangle,
    \label{mean-field:eqn-b}
    \\
    &=
    \mathcal{P}_{N_0} \prod_{p} \Big ( 1 + \eta_p P^\dagger_p \Big ) \vert -; - \rangle,
    \label{mean-field:eqn-c}
    \\
    &=
    \frac{1}{N_0!}\left( \Gamma_\beta^\dagger \right)^{N_0} \vert -;- \rangle
    =
    \vert \Psi_{AGP} (\beta) \rangle,
    \label{mean-field:eqn-d}
  \end{align}
\end{subequations}
where $\eta_p = e^{-\beta \eps_p / 2}$, and we have identified $P^\dagger_p = c^\dagger_p \tilde{c}^\dagger_p$ as the pair-creation operator. As already noted, the un-projected product state in Eq.~\ref{mean-field:eqn-b} is a BCS state and its number-projected version is well known as AGP, with the geminal creation operator $\Gamma_\beta^\dagger$ defined as
\begin{equation}
  \Gamma_\beta^\dagger = \sum_p \eta_p P_p^\dagger.
\end{equation}
Identification of the mean-field state as an AGP is interesting and, with recent developments on efficient evaluation of overlaps and expectation values, as well as geminal based correlated wave function theories,~\cite{khamoshi_efficient_2019, henderson_geminal-based_2019, henderson_correlating_2020, dutta_geminal_2020, khamoshi_correlating_2020} provides a good starting point to include correlation effects.
An improved mean-field description can also be obtained by optimizing both the energy levels $\eps$ and the one-electron basis to find an $H_0$ that minimizes the Helmholtz free energy, in much the same way as Mermin's thermal Hartree Fock theory in Ref.~\onlinecite{mermin_stability_1963}, and as discussed in Refs.~\onlinecite{tanabe_quantum_2005, esashika_effects_2005, nakada_new_2006}.

\subsection{Correlated thermal state}
A plethora of approximate wave function methods are available to study ground-state properties of correlated electronic systems. As we have shown in Refs.~\onlinecite{harsha_thermofield_2019, harsha_thermal_cc_2019}, the thermofield formalism allows for a direct generalization of these methods to finite-temperature.
Since physical electronic systems conserve the number of particles, i.e. $[H, \mathcal{P}_{N_0}] = 0$, we face two options while constructing a correlated approximation to the canonical thermal state: \textit{projection after correlation} (PAC), and \textit{correlation after projection} (CAP).
In PAC, we first construct an approximate grand-canonical thermal state by adding correlation on a broken-symmetry mean-field reference (thermal BCS in our case) and then perform the number-projection,
\begin{equation}
  \vert \Psi \rangle \simeq \mathcal{P}_{N_0} \Omega(\beta) \, \vert 0 (\beta) \rangle; \quad \vert 0 (\beta) \rangle = e^{-\beta H_0 / 2} \vert \mathbb{I} \rangle.
\end{equation}
The correlation operator $\Omega$ is built out of number non-conserving BCS quasiparticles,~\cite{henderson_quasiparticle_2014, signoracci_ab_2015} and the un-projected part of the thermal state, $\Omega (\beta) \vert 0 (\beta) \rangle$, looks like a standard single-reference CI wave function, which simplifies the process of correlating the reference.
In order to carry out the projection efficiently, we use an integral form for the projection operator,~\cite{peierls_collective_1957, bayman_derivation_1960, ring_nuclear_1980} i.e.
\begin{equation}
  \mathcal{P}_{N_0} = \frac{1}{2 \pi} \int_0^{2 \pi} d \phi e^{i \phi (N_0 - N)}.
\end{equation}
Computing matrix elements and overlaps in the presence of $\mathcal{P}$ involves the use of transition density matrices and can be complicated (see e.g., Refs.~\onlinecite{duguet_symmetry_2014, tsuchimochi_communication_2016, duguet_symmetry_2016, qiu_projected_2017, qiu_particle-number_2019})
For CAP, we use the thermal AGP state in Eq.~\ref{mean-field-eqns} as the reference and add correlation using a number-conserving wave operator,
\begin{equation}
  \vert \Psi \rangle \simeq \Lambda(\beta) \vert \Psi_{AGP}(\beta) \rangle
  = \Lambda (\beta) \mathcal{P}_{N_0} \vert 0(\beta) \rangle.
\end{equation}
Contrasting with CAP, the projection problem here is trivial but adding correlation becomes complicated.

Both of these techniques have been explored extensively for ground-state methods.~\cite{degroote_polynomial_2016, tsuchimochi_communication_2016, wahlen-strothman_merging_2017, hermes_combining_2017, qiu_projected_2017, qiu_particle-number_2019, henderson_geminal-based_2019, henderson_correlating_2020, dutta_geminal_2020, khamoshi_correlating_2020} Here, we discuss an example for each: a finite-temperature generalization of the number-projected CI, along the lines discussed by Tsuchimochi \textit{et.~al.} in Ref.~\onlinecite{tsuchimochi_communication_2016}, and an imaginary-time perturbation theory based on the thermal AGP as the reference, as explored in Refs.~\onlinecite{henderson_geminal-based_2019, henderson_correlating_2020, dutta_geminal_2020}

\begin{figure*}[ht]
  \raggedright
  \includegraphics[width=0.49\linewidth]{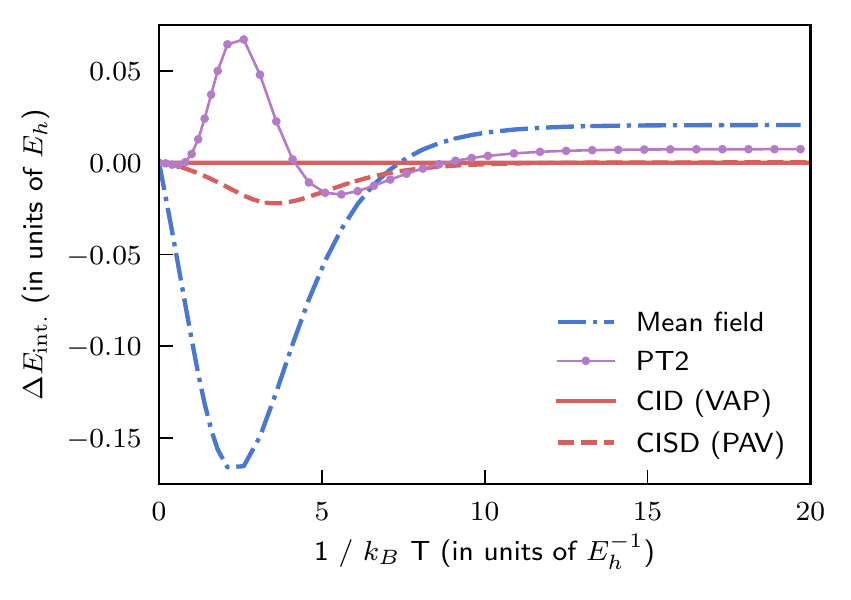}
  \includegraphics[width=0.49\linewidth]{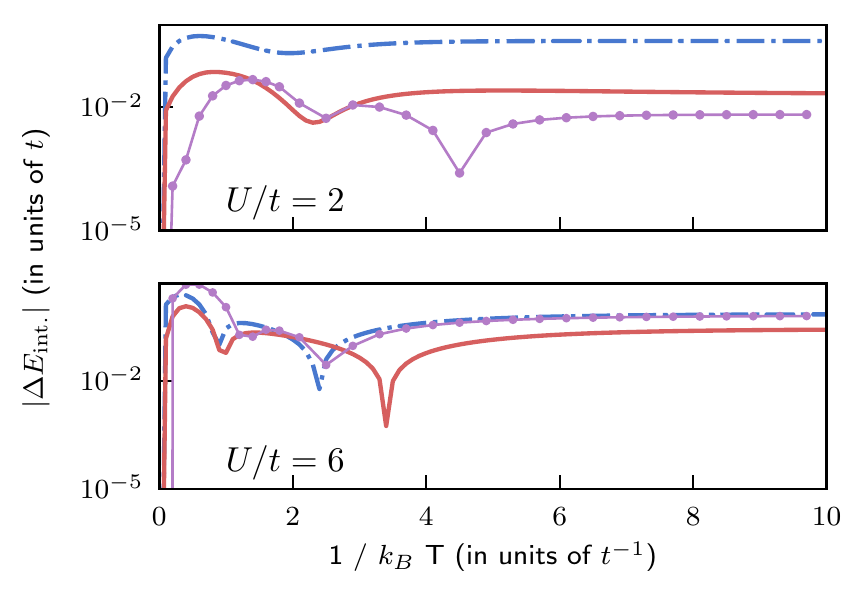}
  \caption{Error in the canonical-ensemble internal energy for (left) the Hydrogen molecule in STO-3G basis with a bond length of $0.74\angstrom$, and (right) the half-filled six-site Hubbard model at $U/t = 2, 6$, computed using the projected BCS wavefunction (mean-field), AGP-based perturbation theory, and projected CI truncated at doubles.}
  \label{fig1}
\end{figure*}

\subsubsection{Projection after correlation}
The number-projected thermal CI state is parameterized as
\begin{equation}
  \vert \Psi_{c} (\beta) \rangle =
  \mathcal{P}_{N_0} e^{t_0} \left ( 1 + T \right ) \vert 0 (\beta) \rangle,
\end{equation}
where $\vert 0 (\beta) \rangle$ is the thermal BCS state at inverse temperature $\beta$, $t_0$ keeps track of the norm of the state (related to the grand potential) and $T$ creates quasiparticle excitations on the BCS,
\begin{equation}
  T = \sum_{pq} t_{pq} a_p^\dagger \tilde{a}_q^\dagger + \frac{1}{4} \sum_{pqrs} t_{pqrs} a^\dagger_p a^\dagger_q \tilde{a}^\dagger_s \tilde{a}^\dagger_r + \ldots.
\end{equation}
The CI amplitudes can be determined in two different ways. One can compute them in the grand-canonical ensemble, as we have done in Ref.~\onlinecite{harsha_thermofield_2019}, and then perform a one-shot projection. This approach is generally known as \textit{projection after variation} (PAV). Alternatively, the amplitudes can be computed in the presence of the projection operator by solving the imaginary-time evolution equation, referred to as \textit{variation after projection} (VAP). VAP allows for more variational freedom and thus, performs better than PAV. Accordingly, we focus our attention on VAP hereafter.

Substituting this CI ansatz into Eq.~\ref{imag-time-evol-eq} and evaluating overlaps of the resulting equation against the ground and excited BCS states, we get
\begin{multline}
  \int_0^{2\pi} d\phi \, \langle 0 (\beta) \vert \nu e^{i \phi (N_0 - N)} \left(
    (1 + T) \frac{d t_0}{d \beta} + \frac{d T}{d \beta}
  \right) \vert 0 (\beta) \rangle
  \\
  = \int_0^{2\pi} d \phi \, \langle 0(\beta) \vert \nu e^{i \phi (N_0 - N)} \bar{H} \vert 0(\beta) \rangle,
  \label{ci-imaginary-time}
\end{multline}
where $\bar{H}$ is the effective Hamiltonian,
\begin{equation}
    \bar{H} =
    -\frac{1}{2} \Big(H (1 + T) - (1 + T) H_0 \Big),
\end{equation}
and $\nu$ takes values from $\{1,~ \tilde{a}_q a_p ,~ \tilde{a}_r \tilde{a}_s a_q a_p , ~\ldots \}$ to construct ground and excited BCS states for the bra.
Both the amplitudes as well as the quasiparticle operators are functions of temperature, therefore the $\beta$-derivative can be broken down into the derivative of the amplitudes and that of the operator parts,
\begin{equation}
  \frac{d T}{d \beta} = \frac{d_{amp} T}{d \beta} + \frac{d_{op} T}{d \beta}.
\end{equation}
We can rewrite Eq.~\ref{ci-imaginary-time} as a system of first-order ODEs that govern the evolution of the CI-amplitudes,
\begin{equation}
  \sum_{\mu} A_{\nu \mu} \cdot \frac{\partial t_\mu}{\partial \beta} = B_\nu,
  \label{cisd-evolution-equation}
\end{equation}
where $A$ is the overlap matrix,
\begin{subequations}
  \begin{align}
    A_{\nu \mu} &=
    \int_0^{2\pi} d\phi \, \langle \nu (\beta) \vert e^{-i \phi (N - N_0)} \mathcal{L}_\mu \vert 0(\beta) \rangle,
    \\
    \mathrm{with} \quad
    \mathcal{L}_\mu &= \begin{cases}
      1 + T, & \mu = 1\\
      \mu, & \mu \in \{a_p^\dagger \tilde{a}_q^\dagger, \, a_p^\dagger a_q^\dagger \tilde{a}_s^\dagger \tilde{a}_r^\dagger \}
    \end{cases}.
  \end{align}
\end{subequations}
The right hand side vector $B_\nu$ is given by
\begin{subequations}
  \begin{align}
    B_\nu  &=
    \int_0^{2\pi} d \phi \, \langle \nu (\beta) \vert e^{-i \phi (N - N_0)} \mathcal{R} \vert 0 (\beta) \rangle,
    \\
    \mathcal{R} &=
    \bar{H} - \frac{d_{op} T}{\partial \beta}.
  \end{align}
\end{subequations}
Here, we have used $\nu, \mu$ as a composite notation for the ground and excited quasiparticle states.
Equation~\ref{cisd-evolution-equation} can be integrated starting from $\beta = 0$, where $T = 0$ is the exact initial condition.

\subsubsection{Correlation after projection}
For correlation after projection method, a numerical integration to perform the projection is not required as it uses a strictly number conserving state, the thermal AGP, as the reference. As an example for this approach, we consider the perturbation theory (PT), where we partition the Hamiltonian as $H = H_0 + \lambda V$, where $H_0$ is the mean-field contribution and $V$ acts as a perturbation. The canonical thermal state can be expanded as a series in $\lambda$,
\begin{subequations}
  \begin{align}
    \vert \Psi(\beta) \rangle
    &= \vert \Psi_0 \rangle + \lambda \vert \Psi_1 \rangle + \lambda^2 \vert \Psi_2 \rangle + \ldots,
    \\
    &= e^{-\beta H_0 / 2} \left( \vert \phi_0 \rangle + \lambda \vert \phi_1 \rangle + \lambda^2 \vert \phi_3 \rangle + \ldots \right).
    \label{pt-series-psi}
  \end{align}
\end{subequations}
Substituting this form for $\vert \Psi \rangle$ in Eq.~\ref{imag-time-evol-eq} and collecting terms at various orders in $\lambda$ gives $\partial \vert \phi_0 \rangle / \partial \tau = 0$, or equivalently $\vert \Psi_0 \rangle = \vert \Psi_{AGP}(\beta) \rangle$ for terms at $\mathcal{O}(\lambda^0)$, and
\begin{equation}
  \frac{\partial}{\partial \tau} \vert \phi_n \rangle
  =
  -\frac{1}{2} e^{\tau H_0 / 2} V e^{-\tau H_0 / 2} \vert \phi_{n-1} \rangle
  \label{pt-eqns}
\end{equation}
for $\mathcal{O}(\lambda^n), \,n \geq 1$. Integrating Eq.~\ref{pt-eqns} yields perturbative corrections identical to those in a time-dependent interaction picture theory. We work in a basis where $H_0$ is diagonal. This allows us to integrate the equations analytically.
Detailed notes on both the projected CI and the AGP-based perturbation theory are available in the Supplemental Information.

\begin{figure*}[ht]
  \raggedright
  \includegraphics[width=0.49\linewidth]{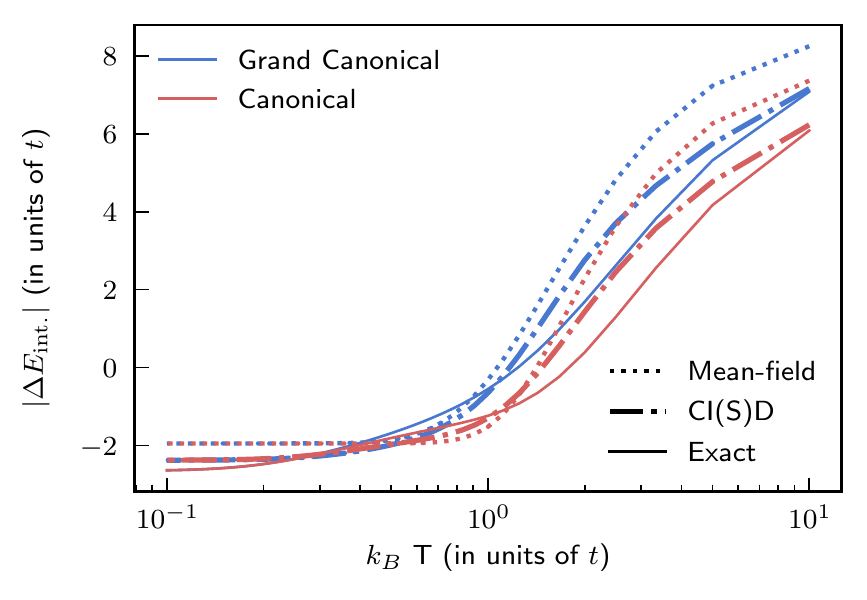}
  \includegraphics[width=0.49\linewidth]{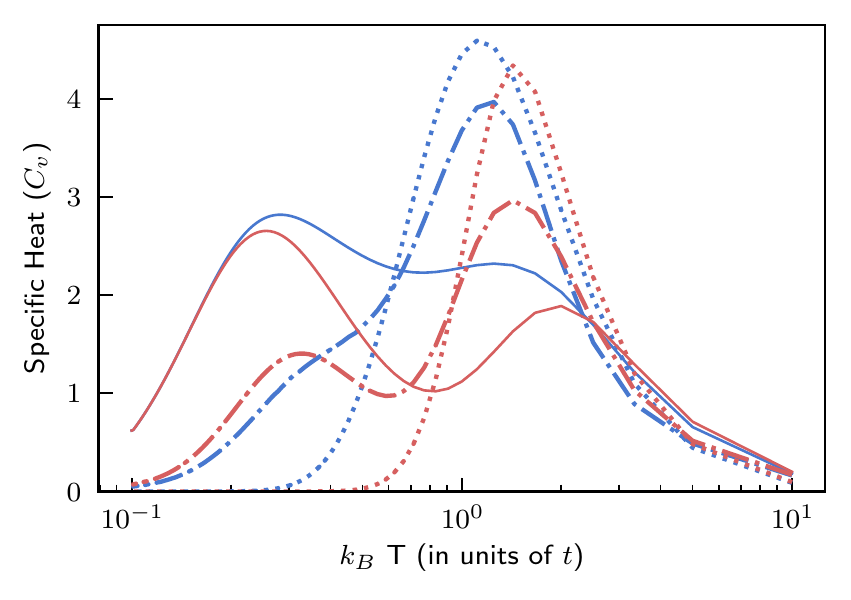}
  \caption{Comparison of total internal energies and specific heats for the half-filled six-site Hubbard model with $U/t=6$ as a function of temperature. The mean-field, CI and exact results highlight the difference between the grand-canonical (blue) and the canonical (red) ensemble properties.}
  \label{fig2:hubbard-cv}
\end{figure*}

\begin{figure*}[t]
  \raggedright
  \includegraphics[width=0.49\linewidth]{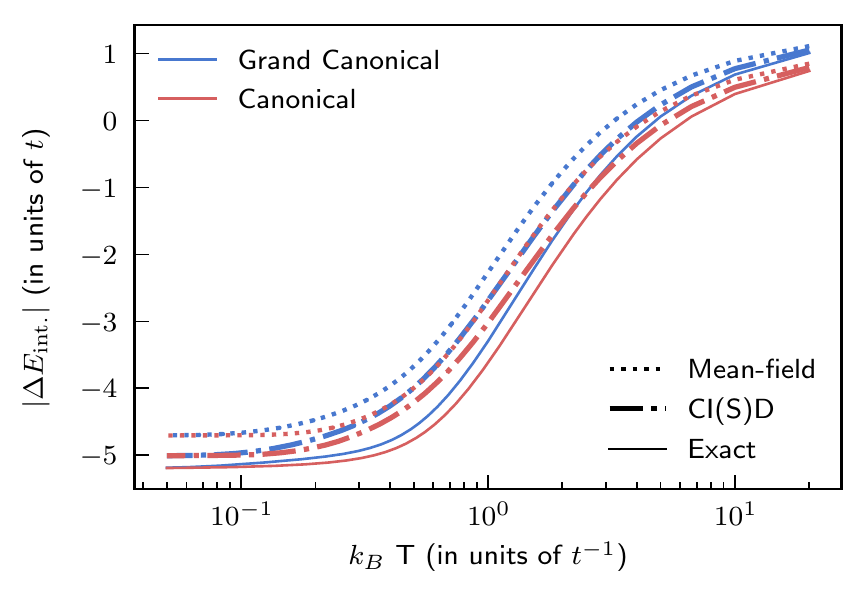}
  \includegraphics[width=0.49\linewidth]{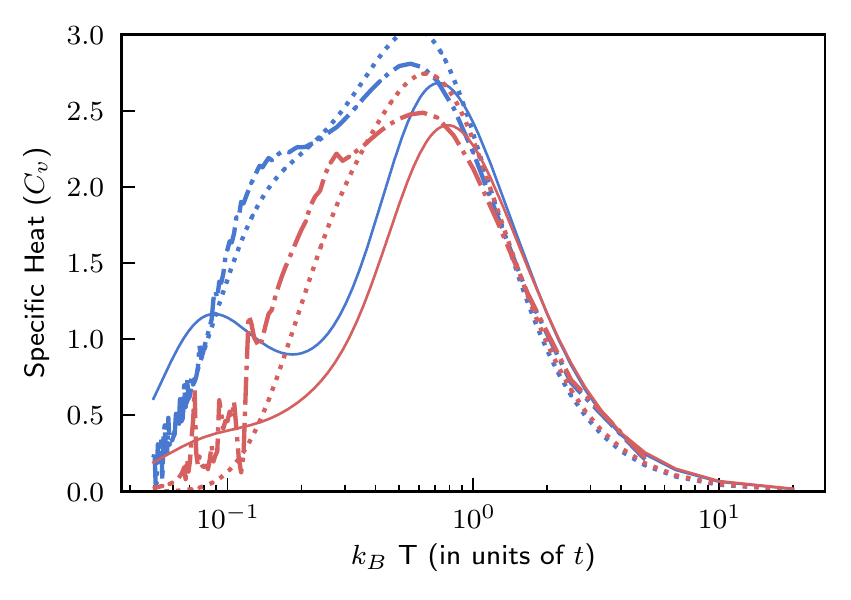}
  \caption{Comparison of total internal energies and specific heats for the six-site Hubbard model with $U/t=4$ and four electrons as a function of temperature. The mean-field, CI and exact results highlight the difference between the grand-canonical (blue) and the canonical (red) ensemble properties.}
  \label{fig3:hubbard-cv}
\end{figure*}

\section{Implementation Details}
We use ground-state Hartree-Fock eigenvalues to build $H_0$, which in turn are used to define the mean-field reference state (thermal BCS for the projected CI, and thermal AGP for the perturbation theory).
We have used PySCF~\cite{pyscf} to generate the Hartree-Fock eigenvalues and Hamiltonian matrix elements.
One can also choose an $H_0$ that optimizes the free energy at any given $\beta$.
While this may lead to a better thermal reference state, it makes the underlying equations very complicated and therefore, in this work, we work with a fixed $H_0$.
This is also analogous to typical interaction picture theories.

To gauge the relevance of optimization, in Fig. 1 of the Supplemental Information, we compare the performance of thermal AGP with optimized and unoptimized $\eta$'s for various benchmark systems which we study below.
We notice that for larger systems, the optimization of $\eta$'s does not introduce any significant improvement, therefore justifying the use of unoptimized $H_0$.

Both the projected CI (Eq.~\ref{cisd-evolution-equation}) and perturbation theory (Eq.~\ref{pt-eqns}) equations are integrated starting from $\beta = 0$, where mean-field is exact and the correct initial condition is known.
The cost for computing these equations is similar to standard projected quasiparticle or AGP-based CI, i.e. $\mathcal{O}(N^6)$.
While the PT2 corrections can be obtained by a straightforward integration of the underlying quantities along the imaginary-time axis, the projected CI amplitudes satisfy a set of linear ordinary differential equations (ODE).
Exact solution of these ODE's requires inversion of the overlap matrix $A$, which is computationally expensive. Moreover, $A$ may also have zero or near-zero eigen modes. To avoid these issues, at each $\beta$-grid point, we solve for the derivative vector iteratively using MinresQLP,~\cite{choi_minres-qlp_2011, choi_algorithm_2014} a robust algorithm for singular linear systems, and then use a fourth order Runge-Kutta method to perform the integration. This adds an additional cost to the projected CI theory.
In all the data presented below, we use a step size of $\Delta \beta = 0.001$ or smaller to integrate the ODE in projected CI, which is sufficiently small to guarantee convergence for the Runge-Kutta method (see Section V in the Supplemental Information).

We also observe that the partial traces of higher rank terms in the CI operator are proportional to the lower-rank terms, e.g. CI with single and double excitations is equivalent to CI with just the double excitations. To avoid linear dependencies in the overlap matrix, we keep only the highest rank terms in our truncated CI theory.

The number projection in the projected CI equations is carried out numerically and converges rapidly as the number of grid points becomes greater than the number of spin-orbitals.

\section{Results}
We apply the projected CI with double excitations (CID), as well as the second order perturbation theory (PT2) to small molecular and model systems to highlight the performance of these finite-temperature canonical ensemble methods against exact benchmark results.
Figure~\ref{fig1} shows error in the canonical-ensemble internal energy for the Hydrogen molecule in the minimal STO-3G basis and at a bond length of $0.74 \angstrom$ (left) and the six-site Hubbard model with $U/t = 2, 6$ (right). The results compare the performance of projected thermal BCS or AGP (which is indicated in the plot by `mean-field'), PAV and VAP projected thermal CISD and CID respectively, and AGP-based PT2.
We use the ground-state spin-restricted Fock operator as our unoptimized $H_0$ for the Hydrogen molecule and the Hubbard model with $U/t = 2$, and the spin-unrestricted Fock operator for $U/t = 6$.
It is apparent that the mean-field approach misses out a lot of correlation, a part of which is recovered by CID and PT2. In fact, the VAP CID, like its ground-state analogue and unlike the grand-canonical CISD in Ref~\onlinecite{harsha_thermofield_2019}, is exact for a two-electron system like the Hydrogen molecule, and expectedly outperforms the PAV approach. The second order perturbation theory, though not exact for the two-electron case, also improves upon the mean-field results.
All the CI and PT results approach their appropriate ground-state counterparts in the zero temperature limit, i.e. the number-projected CID approaches ground-state CISD, and the AGP-based perturbation theory approaches ground-state perturbation theory as $\beta \rightarrow \infty$.
In particular, we notice that the AGP-based PT2 performs better than projected CI for $U/t=2$, but does not introduce any significant improvement over the mean-field for $U/t = 6$. This, in fact, is analogous to the ground-state performance of these theories (see Fig. 2 and Table 1 in the Supplemental Information for the ground-state results).

To highlight the merits of the projected CI theory over mean-field, as well as the distinction between canonical and grand-canonical ensemble properties, we plot the total internal energy (left panel) and the specific heat (right panel) for the six-site Hubbard model with $U/t=6$ at half-filling in Fig.~\ref{fig2:hubbard-cv}.
We compare the mean-field theory, CISD for grand-canonical, and CID for the VAP projected CI against exact numerical results.
We remind the reader that the grand-canonical mean-field state is a thermal BCS which, upon number-projection, gives the canonical thermal state. We use the spin-unrestricted Fock operator to construct $H_0$.
We notice a striking difference in the behavior of the specific heat in the two ensembles. The two different peaks in the exact specific heat curves (shown in solid blue and red lines), which correspond to the spin and charge excitation energy scales, are more pronounced and distinct in the canonical ensemble. While the mean-field theory completely fails to account for the spin-excitation peak, the projected CID performs better both qualitatively and quantitatively.

We repeat this exercise for the hole-doped six-site Hubbard model with $U/t=2$ and four electrons to further demonstrate the difference between the two ensembles. We use the spin-restricted Fock operator to construct $H_0$. The results are plotted in Fig.~\ref{fig3:hubbard-cv}.
Notice that unlike the half-filled case, this hole-doped Hubbard model shows appreciable different results in the canonical and grand-canonical ensembles. This is because the half-filled Hubbard model corresponds to the lowest energy state in Fock space, and excitations to sectors with different particle number are high in energy and are effectively frozen out in the low-temperature limit so that the grand canonical ensemble becomes effectively canonical. This is not the case for the doped Hubbard model.

Finally, we note that the low-temperature specific heat results in Fig.~\ref{fig3:hubbard-cv}, for both the grand-canonical and the canonical CI, are noisy. We attribute this noise to two different sources:
\begin{enumerate}
  \item The evolution of the CI amplitudes is carried out with respect to the inverse temperature $\beta$, and we compute the specific heat as
  \begin{equation}
    C_v = -\beta^2 \frac{d E}{d \beta}.
  \end{equation}
  Any error in the integration due to the finite step size would be amplified by a factor of $\beta^2$. This explains the noise present in both the grand-canonical and the canonical CI.

  \item Recall that for the projected CI, we solve a generalized linear equation (see Eq.~\ref{cisd-evolution-equation}). As we approach low temperatures (or large $\beta$), the number of near-zero modes in the overlap matrix $A$ becomes large, which leads to inconsistencies in the solution, further adding to the noise.
\end{enumerate}

\section{Conclusion}

We have presented a theory to generalize correlated ground-state wave function theories, namely Hartree-Fock, perturbation theory, and CI, to study canonical ensemble thermal properties in fermionic many-body systems. In the low-temperature regime, where the canonical ensemble is most applicable, these methods perform as well as their ground-state counterparts for the benchmark problems studied. The ability to build both canonical and grand-canonical methods also signifies the robustness of thermofield theory for finite-temperature wave function methods. At zero temperature, one is generally required to go to much higher orders in CI or PT to obtain highly accurate results and better alternatives, such as the coupled cluster theory and multi-reference methods, are generally preferred.
While a number-projected formulation of the coupled cluster theory for the ground-state has been worked out in Ref.~\onlinecite{qiu_particle-number_2019}, the underlying equations are complicated for a direct generalization to finite temperatures.
Our work is a first step towards achieving finite-temperature analogues of such sophisticated techniques.
It also establishes a firm standing ground to build number-conserving finite-temperature Monte Carlo methods, something that has been relatively less explored in the QMC community.
Most available thermal methods use an imaginary-time evolution starting from $\beta = 0$ or $T = \infty$, while one is generally interested in low and intermediate temperature scales. A theory that uses ground-state or $T=0$ as the starting point would not only be more practical, but also allow us to systematically eliminate the inconsistencies in the projected CI evolution due to the near-zero modes in the overlap matrix.

\section*{Supplemental Information}
Detailed equations for the projected-CID and AGP-based PT2, along with their derivations, are presented in the Supplemental Information. We also provide additional data comparing the optimized and the unoptimized thermal mean-field, ground-state limits of the thermal methods, and convergence of the Runge-Kutta method with respect to the step-size in the evolution of the projected-CI equations.

\begin{acknowledgements}
This work was supported by the U.S. Department of Energy, Office of Basic Energy Sciences, Computational and Theoretical Chemistry Program under Award No. DE-FG02-09ER16053. G.E.S. acknowledges support as a Welch Foundation Chair (No. C-0036).
\end{acknowledgements}

\section*{Data Availability}
The data that support the findings of this study are available from the corresponding author
upon reasonable request.

\bibliography{CanonTFD}

\end{document}


\title{Supplemental Information for ``Wave function methods for canonical ensemble thermal averages in correlated many-fermion systems''}
\author{Gaurav Harsha}
\affiliation{Department of Physics and Astronomy, Rice University, Houston TX 77005}
\author{Thomas M. Henderson}
\affiliation{Department of Physics and Astronomy, Rice University, Houston TX 77005}
\affiliation{Department of Chemistry, Rice University, Houston TX 77005}
\author{Gustavo E. Scuseria}
\affiliation{Department of Physics and Astronomy, Rice University, Houston TX 77005}
\affiliation{Department of Chemistry, Rice University, Houston TX 77005}

\maketitle

In our manuscript, we presented two theories for building correlated wave function ansatze for computing equilibrium thermal properties in the canonical ensemble: number-projecting a correlated grand-canonical thermal state, and correlating thermal AGP, which is a number-projected mean-field reference. We considered an example of each: number-projected thermal CID and AGP-based perturbation theory. Here, we provide detailed derivations and equations for these methods.
We also provide additional data comparing optimized and unoptimized thermal mean-field, ground-state limit of thermal methods, and convergence of the projected-CI evolution with respect to step-size in the integration.

\section{Number-projected CI}

\subsection{Ansatz}
In the number-projected CI theory, we start with a thermal BCS as our mean-field reference state,
\begin{equation}
  \vert 0(\beta) \rangle = e^{-\beta H_0 / 2} \vert \mathbb{I} \rangle,
  \quad \mathrm{with} \quad
  H_0 = \sum_p \eps_p c_p^\dagger c_p.
\end{equation}
The normalized form $\vert 0 (\beta) \rangle$ becomes,
\begin{equation}
  \vert 0 (\beta) \rangle = \prod_p \left( x_p + y_p c_p^\dagger \tilde{c}_p^\dagger \right) \vert -; - \rangle,
\end{equation}
where $x$ and $y$ coefficients are defined as,
\begin{subequations}
    \begin{align}
        x_p &= \frac{1}{\sqrt{1 + e^{- \beta \epsilon_p}}}, \\
        y_p &= \frac{ e^{- \beta \epsilon_p/2} }{ \sqrt{1 + e^{- \beta \epsilon_p}} }.
    \end{align}
\end{subequations}
The thermal state is then parameterized as a number-projected quasiparticle CI doubles (CID) state,
\begin{subequations}
  \begin{align}
    \vert \Psi \rangle &= \mathcal{P} e^{t_0}(1 + T) \vert 0 (\beta) \rangle,
    \\
    T &= \frac{1}{4} \sum_{pqrs} t_{pqrs} a^\dagger_p a^\dagger_q \tilde{a}^\dagger_s \tilde{a}^\dagger_r.
  \end{align}
\end{subequations}
where $T$ creates double quasiparticle excitations on the reference.
Here, $a, \tilde{a}$ are quasiparticle operators that annihilate the mean-field reference, i.e. $a_p \vert 0 (\beta) \rangle = 0 = \tilde{a}_p \vert 0(\beta) \rangle$, and are related to the physical creation/annihilation operators via a BCS transformation,
\begin{equation}
    \begin{bmatrix}
        a_p  \\
        \tilde{a}_p^\dagger
    \end{bmatrix} =
    \begin{bmatrix}
        x_p & -y_p\\
        y_p & x_p
    \end{bmatrix}
    \begin{bmatrix}
        c_p\\
        \tilde{c}_p^\dagger
    \end{bmatrix},
    \label{hfb-transformation}
\end{equation}
It is convenient to express the Hamiltonian in terms of the quasiparticle creation/annihilation operators,
\begin{align}
  H &= h_0 + h^{(20)}_{pq} a^\dagger_p a_q + h^{(02)}_{pq} \tilde{a}^\dagger_p \tilde{a}_q
  + h^{(11)}_{pq} (a^\dagger_p \tilde{a}^\dagger_q + \mathrm{h. c.})
  + h^{(31)}_{pqrs} (a^\dagger_p a^\dagger_q \tilde{a}^\dagger_r a_s + \mathrm{h.c.})
  \nonumber
  \\
  & \quad
  + h^{(13)}_{pqrs} (a^\dagger_p \tilde{a}^\dagger_q \tilde{a}^\dagger_r \tilde{a}_s + \mathrm{h.c.})
  + h^{(221)}_{pqrs} (a^\dagger_p a^\dagger_q \tilde{a}^\dagger_s \tilde{a}^\dagger_r + \mathrm{h.c.})
  + h^{(222)}_{pqrs} a^\dagger_p \tilde{a}^\dagger_q \tilde{a}_s a_r.
  \label{Heff}
\end{align}
where we have assumed Einstein summation convention. The matrix elements have been worked out in the Appendix of Ref.~\onlinecite{harsha_thermal_cc_2019}. Lastly, we note that we employ an integral representation for the number-projection operator,
\begin{equation}
  \proj = \frac{1}{2 \pi} \int_0^{2 \pi} d \phi \, e^{i \phi (N_0 - N)},
\end{equation}
where $N$ is the physical number operator and $N_0$ is the target number of particles.

\subsection{CID equations}
Substituting the number-projected CID ansatz in to the imaginary-time evolution equation, and using the fact that the Hamiltonian commutes with the number projection operator $\proj$, we get
\begin{equation}
  \proj \left[
    \frac{d t_0}{d \beta} (1 + T)
    + \frac{d T}{d \beta}
  \right] \vert 0(\beta) \rangle
  = -\frac{1}{2} \proj \left[
    H (1 + T) - (1 + T) H_0
  \right] \vert 0(\beta) \rangle
\end{equation}
We have noted in the main text (as well as in our recent work in Refs.~\onlinecite{harsha_thermofield_2019, harsha_thermal_cc_2019}) that the derivative of the CI operator splits into the derivative of the amplitudes and that of the temperature-dependent quasiparticle creation operators,
\begin{equation}
  \frac{d T}{d \beta} = \frac{d_{amp} T}{d \beta} +
  \frac{d_{op} T}{d \beta},
\end{equation}
where the operator derivative can be worked out directly from the BCS transformation.
Using the above relations and expanding out the integral form of $\proj$, the time-evolution equation simplifies as
\begin{equation}
  \int_0^{2\pi} d \phi \, g(\phi) \, e^{-i \phi N} \, \mathcal{L} \, \vert 0 \rangle
  = \int_0^{2\pi} d \phi \, g(\phi) \, e^{-i \phi N} \, \mathcal{R} \, \vert 0 \rangle,
  \label{proj-state-eq}
\end{equation}
where $g(\phi)$ is the weight of integration, $\mathcal{L}$ and $\mathcal{R}$ are effective LHS and RHS kernels respectively, each defined as
\begin{subequations}
  \begin{align}
    g(\phi) &=
    e^{i \phi N_0} / 2 \pi,
    \\
    \mathcal{L} &=
    \frac{d t_0}{d \beta} (1 + T)
    + \frac{d_{amp} T}{d \beta},
    \\
    \mathcal{R} &=
    - \frac{1}{2} \left[ H (1 + T) - (1 + T) H_0 \right]
    - \frac{d_{op} T}{d \beta}.
  \end{align}
\end{subequations}
The RHS kernel is effectively a four-body operator, and we find it convenient to re-write it in terms of antisymmetrized matrix elements,
\begin{equation}
  \mathcal{R} =
  R_0 + R_{pq} a^\dagger_p \tilde{a}^\dagger_q
  + R_{[pq][rs]} a^\dagger_p a^\dagger_q \tilde{a}^\dagger_s \tilde{a}^\dagger_r
  + R_{[pqr][ijk]} a^\dagger_p a^\dagger_q a^\dagger_r \tilde{a}^\dagger_k \tilde{a}^\dagger_j \tilde{a}^\dagger_i
  + R_{[pqrs][ijkl]} a^\dagger_p a^\dagger_q a^\dagger_r a^\dagger_s \tilde{a}^\dagger_l \tilde{a}^\dagger_k \tilde{a}^\dagger_j \tilde{a}^\dagger_i,
\end{equation}
where we have assumed Einstein summation, and $[\ldots]$ denotes antisymmetrized indices. The matrix elements of $\mathcal{R}$ are defined as,
\begin{subequations}
  \begin{align}
    R_0 &=
    -\frac{1}{2} \left(
      h_0 - \sum_a \eps_a y_a^2
      + \sum_{abcd} h^{(221)}_{abcd} t_{abcd}
    \right),
    \\
    R_{ab} &=
    \frac{1}{2} \left(
      \delta_{ab} \eps_a x_a y_a - h^{(11)}_{ab}
      + \sum_c \eps_c x_c y_c t_{acbc}
      - \sum_{cd} h^{(11)}_{cd} t_{acbd}
      + \sum_{cdi} \left[ h^{(13)}_{cdib} t_{acdi} - h^{(31)}_{cdia} t_{cdbi} \right]
    \right),
    \\
    R_{abcd} &= \frac{1}{4} \chi_{(ab - ba)(cd - dc)} = \frac{1}{4} \left( \chi_{abcd} - \chi_{abdc} - \chi_{bacd} + \chi_{badc} \right),
    \\
    \chi_{abcd} &=
    \frac{1}{2} \left(
      - \frac{1}{4}\left[
        h_0 t_{abcd} + 4 h^{(221)}_{abcd}
      \right]
      + \frac{1}{4} \sum_{i} \left[
        \eps_i y_i^2 t_{abcd} + 2 h^{(02)}_{ci} t_{abdi} + 2 h^{(20)}_{ai} t_{bicd}
      \right]
    \right.
    \nonumber
    \\
    & \quad \quad
    \left.
      - \frac{1}{2} \sum_{ij} \left[
        t_{abij} h^{(04)}_{ijcd} + 2 h^{(222)}_{acij} t_{bidj} + h^{(40)}_{abij} t_{ijcd}
      \right]
    \right),
    \\
    R_{pqrabc} &= \frac{1}{9} \chi_{(pqr - qpr - rqp)(abc - bac - cba)},
    \\
    \chi_{pqrabc} &=
    \frac{1}{2} \left(
      \frac{1}{4} \delta_{ap} \eps_p x_p y_p t_{qrbc} - \frac{1}{4} h^{(11)}_{pa} t_{qrbc}
    \right)
    + \frac{1}{4} \sum_i \left( h^{(13)}_{pbci} t_{qrai} + h^{(31)}_{qrai} t_{ipbc} \right),
    \\
    R_{pqrsabcd} &= \frac{1}{36} \chi_{(pqrs - prqs - rqps - sqrp + rspq + psqr)(abcd - acbd - cbad - dbca + cdab + adbc)},
    \\
    \chi_{pqrsabcd} &=
    -\frac{1}{8} h^{(221)}_{pqab} t_{rscd}.
  \end{align}
\end{subequations}
Ultimately, we get the equations governing the $\beta$-evolution of the CI amplitudes by taking the expectation value of Eq.~\ref{proj-state-eq} against ground and doubly-excited quasiparticle subspace on the bra state, i.e.
\begin{equation}
  \int d \phi \, g(\phi) \, \langle 0 \vert \nu e^{-i \phi N} \, \mathcal{L} \, \vert 0 \rangle
  = \int d \phi \, g(\phi) \, \langle 0 \vert \nu  e^{-i \phi N} \, \mathcal{R} \, \vert 0 \rangle,
  \label{contracted-eq-prelim}
\end{equation}
where $\nu = \{ 1, \, \tilde{a}_r \tilde{a}_s a_q a_p \}$.
After some rearrangements, the working equation takes the form
\begin{equation}
  \int d \phi \, w(\phi) \, \langle \phi \vert \bar{\nu} \, \mathcal{L} \, \vert 0 \rangle
  = \int d \phi \, w(\phi) \, \langle \phi \vert \bar{\nu} \, \mathcal{R} \, \vert 0 \rangle,
  \label{contracted-eq-final}
\end{equation}
where
\begin{subequations}
  \begin{align}
    \bar{\nu} &= e^{i \phi N} \, \nu \, e^{-i \phi N},
    \\
    w(\phi) &= \frac{1}{2 \pi} e^{i \phi N_0} \langle 0 \vert e^{-i \phi N} \vert 0 \rangle,
    \\
    \langle \phi \vert &= \frac{
      \langle 0 \vert e^{-i \phi N}
    }{
      \langle 0 \vert e^{-i \phi N} \vert 0 \rangle
    }.
  \end{align}
\end{subequations}
The overlap in $w(\phi)$ is straightforward to compute:
\begin{equation}
  \langle 0 \vert e^{-i \phi N} \vert 0 \rangle
  = \prod_p \left( x_p^2 + y_p^2 e^{-i \phi} \right).
\end{equation}
The construction of the similarity transformed operator $\bar{\nu}$ can be simplified by introducing rotated quasi-particle operators,
\begin{equation}
  b_p = e^{i \phi N} a_p e^{-i \phi N},
  \quad \mathrm{and,} \quad
  \tilde{b}_p = e^{i \phi N} \tilde{a}_p e^{-i \phi N},
\end{equation}
which gives $\bar{\nu} = \{ 1, \, \tilde{b}_r \tilde{b}_s b_q b_p \}$. The overlaps in Eq.~\ref{contracted-eq-final} can then be evaluated by using a generalized version of Wick's theorem, with the relevant contractions given by,
\begin{subequations}
  \begin{align}
    \langle \phi \vert a^\dagger_p \tilde{a}^\dagger_q \vert 0 \rangle
    &= \delta_{pq} G_p =
    - \langle \phi \vert \tilde{a}^\dagger_p a^\dagger_q \vert 0 \rangle
    =
    \langle \phi \vert \tilde{b}_p b_q \vert 0 \rangle,
    \\
    \langle \phi \vert a_p a^\dagger_q \vert 0 \rangle
    &= \delta_{pq}
    =
    \langle \phi \vert \tilde{a}_p \tilde{a}^\dagger_q \vert 0 \rangle,
    \\
    \langle \phi \vert b_p a^\dagger_q \vert 0 \rangle
    &= \delta_{pq} A_p,
    \\
    \langle \phi \vert \tilde{b}_p \tilde{a}^\dagger_q \vert 0 \rangle
    &= \delta_{pq} B_p,
  \end{align}
\end{subequations}
where we have
\begin{subequations}
  \begin{align}
    A_p &=
    \frac{1}{x_p^2 e^{i \phi} + y_p^2},
    \\
    B_p &=
    \frac{1}{x_p^2 + y_p^2 e^{-i \phi}},
    \\
    G_p &= \frac{x_p y_p (e^{-i \phi} - 1)}{x_p^2 + y_p^2 e^{-i \phi}}.
  \end{align}
\end{subequations}
After all the manipulation, the overlaps for the LHS and the RHS kernels turn out as follows:
\begin{enumerate}
  \item First, for $\bar{\nu} = 1$,
  \begin{subequations}
    \begin{align}
      \langle \phi \vert \mathcal{R} \vert 0 \rangle
      &= R_0 + \sum_a G_a R_{aa} + 2 \sum_{ab} G_a G_b R_{abab} + 6 \sum_{abc} G_a G_b G_c R^{(30)}_{abc}
      + 24 \sum_{abcd} G_a G_b G_c G_d R^{(40)}_{abcd},
      \label{rhs-eq1}
      \\
      \langle \phi \vert \mathcal{L} \vert 0 \rangle
      &= \left( 1 + \frac{1}{2}\sum_{ij} G_i G_j t_{ijij} \right) \frac{d t_0}{d \beta}
      + \frac{1}{2} \sum_{ij} G_i G_j \frac{d t_{ijij}}{d \beta}
      \label{lhs-eq1}
    \end{align}
  \end{subequations}
  where $R^{(30)}_{abc} = R_{abcabc}$ and $R^{(40)}_{abcd} = R_{abcdabcd}$.

  \item And for $\bar{\nu} = \tilde{b}_c \tilde{b}_d b_b b_a$,
  \begin{subequations}
    \begin{align}
      \chi_{ab} &= A_a B_b \left(
        R_{ab} + 4 \sum_i G_i R_{aibi} + 18 \sum_{ij} G_i G_j R^{(31)}_{abij}
        + 96 \sum_{ijk} G_i G_j G_k R^{(41)}_{abijk}
      \right),
      \\
      \langle \phi \vert \tilde{b}_c \tilde{b}_d b_b b_a \mathcal{R} \vert 0 \rangle
      &= \left( 1 - P_{ab} \right) \left( 1 - P_{cd} \right)
      \left[
        \frac{1}{2} \delta_{ac} \delta_{bd} G_a G_b \langle \phi \vert \mathcal{R} \vert 0 \rangle
        + \delta_{ac} G_a \chi_{bd}
      \right.
      \nonumber
      \\
      & \quad
      \left.
        + A_a A_b B_c B_d \left(
          R_{abcd} + 9 \sum_i G_i R^{(32)}_{abcdi} + 72 \sum_{ij} G_i G_j R^{(42)}_{abcdij}
        \right)
      \right],
      \label{rhs-eq3}
      \\
      \zeta_{ab} &= A_a B_b \left[
        \left( \sum_i G_i t_{aibi} \right) \frac{d t_0}{d \beta}
        + \sum_i G_i \frac{d t_{aibi}}{d \beta}
      \right]
      \\
      \langle \phi \vert \tilde{b}_c \tilde{b}_d b_b b_a \mathcal{L} \vert 0 \rangle
      &= \left( 1 - P_{ab} \right) \left( 1 - P_{cd} \right)
      \left[
        \frac{1}{2} \delta_{ac} \delta_{bd} S_a S_b \langle \phi \vert \mathcal{L} \vert 0 \rangle
        + \delta_{ac} S_a \zeta_{bd}
      \right.
      \nonumber
      \\
      &
      \quad \quad
      \left.
        + \frac{A_a A_b B_c B_d}{4} \left(
          t_{abcd} \frac{d t_0}{d \beta} + \frac{d t_{abcd}}{d \beta}
        \right)
      \right],
      \label{lhs-eq3}
    \end{align}
  \end{subequations}
  where $R^{(31)}_{abij} = R_{aijbij}$, $R^{(41)}_{abijk} = R_{aijkbijk}$, $R^{(32)}_{abcdi} = R_{abicdi}$, $R^{(42)}_{abcdij} = R_{abijcdij}$, and $P_{ab}$ is the conventional exchange operator.
\end{enumerate}

\section{Canonical ensemble perturbation theory}
In contrast to the number-projected CI, the perturbation theory adds corrections to the thermal AGP, which is an eigenstate of the total number operator. Moreover, the corrections are also introduced in a number-conserving manner, and therefore an explicit number projection is not required. We first partition the Hamiltonian into a mean-field and an interaction part,
\begin{subequations}
  \begin{align}
    H &= H_0 + \lambda V,
    \\
    H_0 &= \sum_p \eps_p c_p^\dagger c_p,
    \\
    V &= \sum_{pq} v_{pq} c_p^\dagger c_q + \frac{1}{4} \sum_{pqrs} v_{pqrs} c_p^\dagger c_q^\dagger c_s c_r.
  \end{align}
\end{subequations}
The canonical-ensemble thermal state, which obeys the imaginary-time evolution equation $d_\tau \vert \Psi (\tau) \rangle = -H \vert \Psi (\tau) \rangle / 2$, can be expressed as a perturbation series expansion,
\begin{equation}
  \vert \Psi (\tau) \rangle = \vert \Psi_0 \rangle + \lambda \vert \Psi_1 \rangle + \lambda^2 \vert \Psi_2 \rangle + \ldots.
\end{equation}
Substituting this form of $\vert \Psi (\tau) \rangle$ into the imaginary-time evolution equation, we get
\begin{equation}
  d_\tau \left(
    \vert \Psi_0 \rangle + \lambda \vert \Psi_1 \rangle + \lambda^2 \vert \Psi_2 \rangle + \ldots
  \right)
  =
  -\frac{1}{2} (H_0 + \lambda V) \left(
    \vert \Psi_0 \rangle + \lambda \vert \Psi_1 \rangle + \lambda^2 \vert \Psi_2 \rangle + \ldots
  \right).
\end{equation}
Equating the terms at various orders in $\lambda$ on the left and right hand sides gives
\begin{subequations}
  \label{pt-lambda-orders}
  \begin{align}
    \mathcal{O}(\lambda^0): \quad \quad \quad \quad
    d_\tau \vert \Psi_0 \rangle
    &=
    -\frac{1}{2} H_0 \vert \Psi_0 \rangle,
    \\
    \mathcal{O}(\lambda^1): \quad \quad \quad \quad
    d_\tau \vert \Psi_1 \rangle
    &=
    -\frac{1}{2} \big( H_0 \vert \Psi_1 \rangle + V \vert \Psi_0 \rangle \big),
    \\
    \mathcal{O}(\lambda^2): \quad \quad \quad \quad
    d_\tau \vert \Psi_2 \rangle
    &=
    -\frac{1}{2} \big( H_0 \vert \Psi_2 \rangle + V \vert \Psi_1 \rangle \big),
  \end{align}
\end{subequations}
and similarly for higher orders. For the purpose of this work, we confine ourselves to the second order perturbation theory. Without the loss of any generality, we can assume $\vert \Psi_n \rangle = e^{-\tau H_0/2} \vert \phi_n \rangle$. This simplifies Eq.~\ref{pt-lambda-orders} to,
\begin{subequations}
  \label{pt-lambda-orders-2}
  \begin{align}
    d_\tau \vert \phi_0 \rangle &= 0,
    \\
    d_\tau \vert \phi_1 \rangle
    &=
    -\frac{1}{2} \, e^{\tau H_0 / 2} V e^{-\tau H_0 / 2} \vert \phi_0 \rangle,
    \\
    d_\tau \vert \phi_2 \rangle
    &=
    -\frac{1}{2} \, e^{\tau H_0 / 2} V e^{-\tau H_0 / 2} \vert \phi_1 \rangle.
  \end{align}
\end{subequations}
These equations can then be integrated, starting from a known initial condition (generally $\tau = 0$), to the desired inverse temperature $\beta$.
Recognizing that at $\beta = 0$, extreme AGP is the exact canonical thermal state, we get
\begin{equation}
  \vert \phi_0 (\beta) \rangle = \vert AGP (\beta=0) \rangle = \left( \sum_p P^\dagger_p \right)^{N_0} \vert -; -\rangle,
\end{equation}
where $N_0$ is again the desired number of particles and $P^\dagger_p = c_p^\dagger \tilde{c}_p^\dagger$. Consequently, we recover the thermal AGP as the zeroth order approximation,
\begin{equation}
  \vert \Psi_0 (\beta) \rangle = e^{-\beta H_0 / 2} \vert \textrm{AGP}(0) \rangle = \vert \textrm{AGP}(\beta) \rangle.
\end{equation}
At the first order, we have
\begin{subequations}
  \begin{align}
    \vert \Psi_1(\beta) \rangle
    &= -\frac{e^{-\beta H_0 / 2}}{2} \int_0^{\tau} d \tau \, e^{\tau H_0 / 2} V e^{-\tau H_0 / 2} \vert \textrm{AGP}(0) \rangle,
    \\
    &= -\frac{1}{2} \int_0^{\beta} d \tau \, e^{-(\beta - \tau) H_0 / 2} V e^{(\beta -\tau) H_0 / 2} \vert \Psi_0(\beta) \rangle,
    \\
    &= -\frac{1}{2} \int_0^{\beta} d \tau \, e^{-\tau H_0 / 2} V e^{\tau H_0 / 2} \vert \Psi_0 (\beta) \rangle,
  \end{align}
\end{subequations}
Likewise, the second order contribution to the wave function is
\begin{subequations}
  \begin{align}
    \vert \Psi_2 (\beta) \rangle
    &= -\frac{e^{-\beta H_0 / 2}}{2} \int_0^{\beta} d \tau \, e^{\tau H_0 / 2} V e^{-\tau H_0 / 2} \vert \phi_1(\tau) \rangle,
    \\
    &= \frac{e^{-\beta H_0 / 2}}{4} \int_0^{\beta} d \tau \, e^{\tau H_0 / 2} V e^{-\tau H_0 / 2}
    \int_{0}^{\tau} d \tau' \, e^{\tau' H_0 / 2} V e^{-\tau' H_0 / 2} \vert \phi_0 \rangle,
    \\
    &= \frac{1}{4} \int_0^{\beta} d\tau \, e^{-(\beta - \tau) H_0 / 2} V e^{(\beta - \tau) H_0 / 2}
    \int_0^{\tau} d \tau' \, e^{-(\beta - \tau') H_0 / 2} V e^{(\beta - \tau') H_0 / 2} \vert \Psi_0 (\beta) \rangle,
    \\
    &= \frac{1}{4} \int_0^\beta d \tau \, e^{-\tau H_0 / 2} V e^{\tau H_0 / 2}
    \int_{0}^{\beta - \tau} d \tau' \, e^{-(\beta - \tau') H_0 / 2} V e^{(\beta - \tau') H_0 / 2} \vert \Psi_0 (\beta) \rangle,
    \\
    &= \frac{1}{4} \int_0^\beta d \tau \, e^{-\tau H_0 / 2} V e^{\tau H_0 / 2}
    \int_\tau^\tau d \tau' \, e^{-\tau' H_0 / 2} V e^{\tau' H_0 / 2} \vert \Psi_0 (\beta) \rangle.
  \end{align}
\end{subequations}
These perturbative corrections are analogous to the Dyson series expansion for interaction picture, imaginary-time perturbation theory.
The diagonal form of $H_0$ allows us to perform the similarity transformation of $V$ easily,
\begin{equation}
  e^{-\tau H_0 / 2} V e^{\tau H_0 / 2} = v_{pq} e^{-\tau \Delta_{pq} / 2} p^\dagger q + \frac{1}{4} v_{pqrs} e^{-\tau \Delta_{pqrs} / 2} p^\dagger q^\dagger s r,
\end{equation}
where $\Delta_{pq} = \eps_p - \eps_q$, and $\Delta_{pqrs} = \eps_p + \eps_q - \eps_r - \eps_s$. The resulting equations can then be integrated analytically.

\begin{figure}[t]
  \centering
  \includegraphics[width=0.47\linewidth]{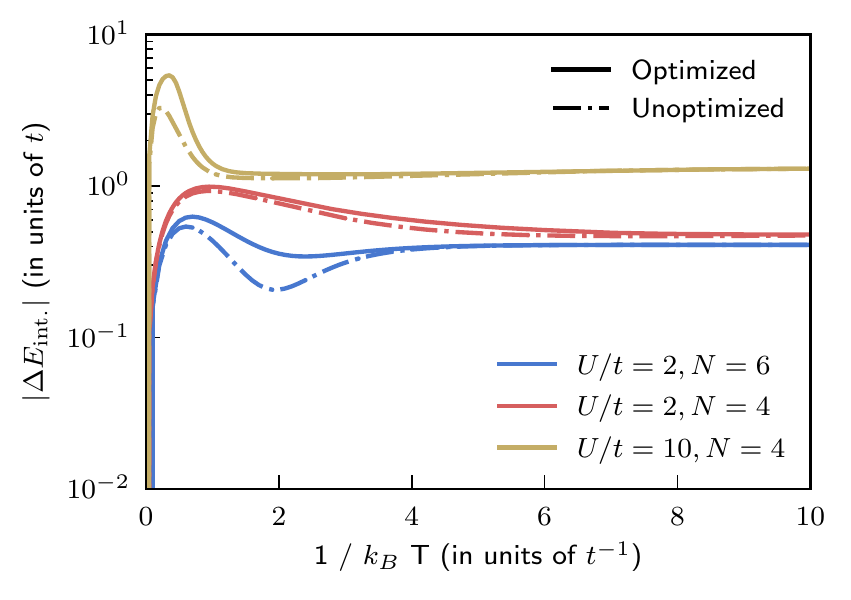}
  \includegraphics[width=0.47\linewidth]{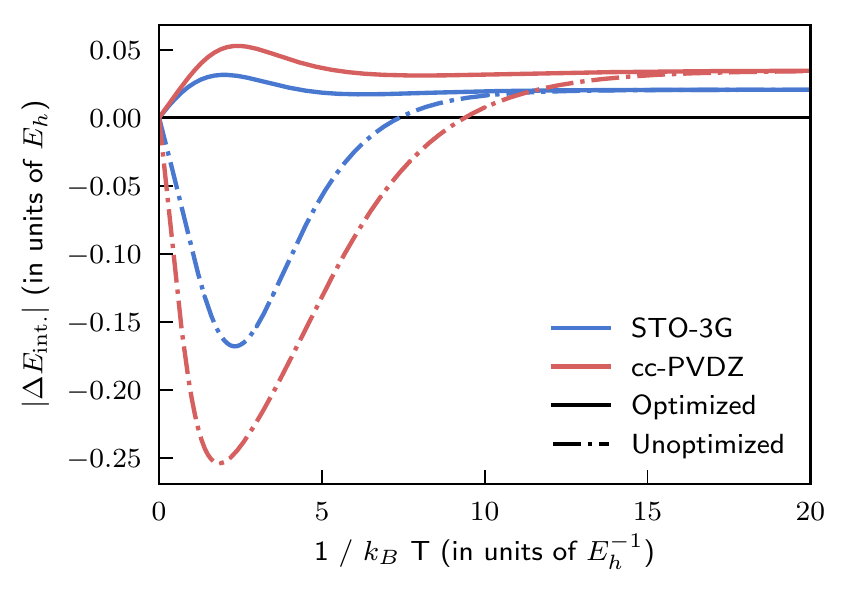}
  \caption{Error in the canonical-ensemble internal energy for the 6-site Hubbard model with $U/t = 2$ and $10$ at different filling fractions (left panel), and the Hydrogen molecule at bond-length of $0.74\AA$ in STO-3G and cc-PVDZ bases (right panel). The data compares mean-field thermal AGP results with optimized versus unoptimized $\eta$-parameters.}
  \label{figSI1}
\end{figure}

\begin{figure}[b]
  \centering
  \includegraphics[width=0.47\linewidth]{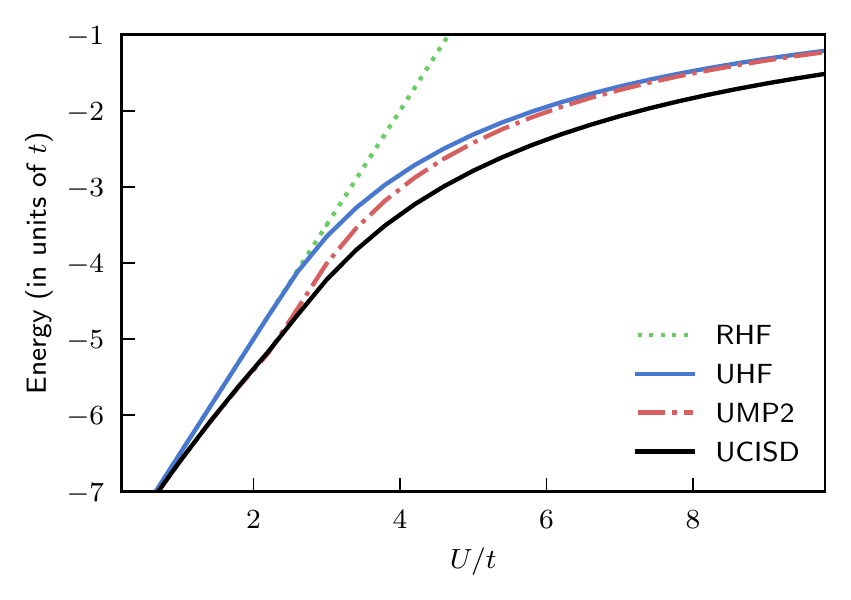}
  \caption{Ground-state energy for the six-site Hubbard model at various $U/t$ values. The data compares RHF, UHF, UMP2, and UCISD approximations to the ground-state energy.}
  \label{figSI2}
\end{figure}

\section{Comparison of optimized and unoptimized AGP}
In this section, we compare the performance of thermal mean-field results obtained using optimized and unoptimized $\eta$'s in the thermal AGP. The left panel in Fig.~\ref{figSI1} presents the data for the 6-site Hubbard model with $U/t = 2$ and $N=4, 6$, as well as $U/t = 10$ and $N=4$. Except in the case of very high temperatures, or very low $\beta$, the optimized and the unoptimized mean-field theories perform more or less similarly. The right panel in Fig.~\ref{figSI1} shows similar data for Hydrogen molecule at a bond-length of $0.74 \AA$ in two different basis sets. In contrast to the Hubbard model results, the difference between the optimized and the unoptimized mean-field theories is more pronounced for Hydrogen. However, as we have noted in the main text, number-projected CID is exact for a two-electron system whether or not we start with a better reference.
This allows us to safely use unoptimized mean-field as a reference to construct correlated theories such as projected CID and AGP-based perturbation theory.
We would like to note that these results do not take orbital optimization into account, but instead just minimize the free energy with respect to the BCS/AGP parameters.

\section{Ground-state results for the Hubbard model}
In this section, we present spin-restricted and unrestricted Hartree-Fock (RHF and UHF), unrestricted second order perturbation theory (UMP2), and unrestricted CISD results for the ground-state energy in the six-site Hubbard model at various $U/t$ values (see Fig.~\ref{figSI2}). We used PySCF~\cite{pyscf} to compute these results.
An important observation is that as we increase the correlation strength $U/t$, the improvement in the second-order unrestricted perturbation theory (UMP2) energy over the UHF becomes smaller and smaller.

Performance of thermal perturbation theory, presented in the main text, is consistent with this observation.
In Table~\ref{tableSupp}, we compare finite-temperature internal energies at the largest $\beta$-value used in the main text with the corresponding ground-state theories. The results demonstrate the convergence of the finite-temperature theories to their ground-state counterparts.

\begin{table}[h]
  \caption{\label{tableSupp}Ground-state limit of finite-temperature wave function theories. All the energies are reported in atomic units have been rounded off to six significant digits, and all the Hubbard models are 6-site systems.}
    \begin{tabular}{|p{4cm}|p{1cm}|p{1.5cm}|p{1.5cm}|p{1.5cm}|p{1.5cm}|p{1.5cm}|p{1.5cm}|}
      \hline
      \multirow{2}{*}{System} & \multirow{2}{*}{$\beta$} & \multicolumn{2}{c|}{Mean-field} & \multicolumn{2}{c|}{CISD} & \multicolumn{2}{c|}{PT2} \\
      \cline{3-8} & & Thermal & Ground-state & Thermal & Ground-state  & Thermal  & Ground-state \\
      \hline

      Hydrogen (STO-3G) & 20 & -1.11676 & -1.11676 & -1.13727 & -1.13728 & -1.12990 & -1.12990
      \\

      Hubbard, $U/t=2$, $N=6$ & 10 & -5.0 & -5.0 & -5.38753 & -5.38878 & -5.40278 & -5.40278
      \\

      Hubbard, $U/t=2$, $N=4$ & 10 & -4.70295 & -4.70296 & -5.00657 & -5.00685 & -5.08802 & -5.08805
      \\

      Hubbard, $U/t=6$, $N=6$ & 10 & -1.94760 & -1.94760 & -2.37682 & -2.37822 & -2.01540 & -2.01540
      \\
      \hline
    \end{tabular}
\end{table}

\begin{figure}[t]
  \centering
  \includegraphics[width=0.47\linewidth]{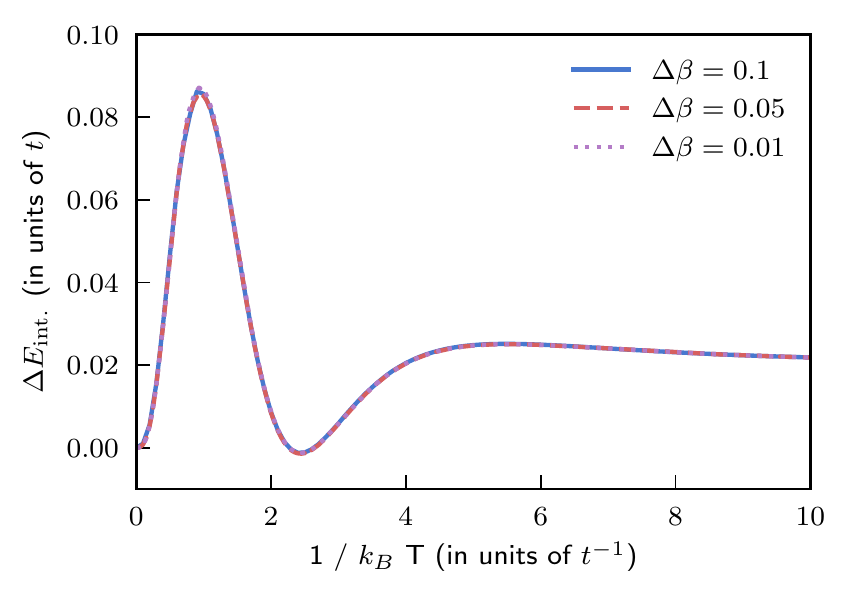}
  \includegraphics[width=0.47\linewidth]{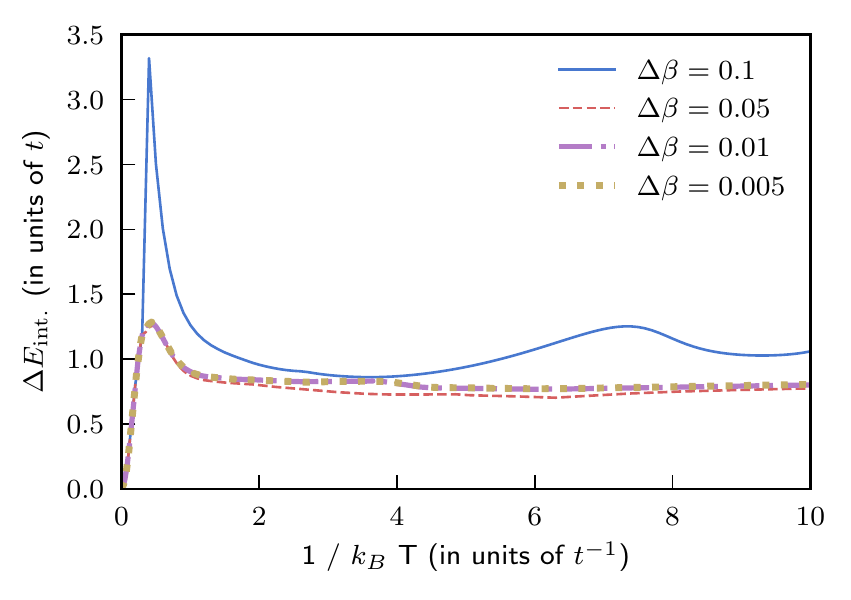}
  \caption{Error in the projected CI internal energy for the six-site Hubbard model with $U/t = 2$ and six-electrons (left panel), and $U/t = 10$ and four electrons (right panel), with various step-sizes used to integrate the projected-CI ODEs.}
  \label{figSI3}
\end{figure}

\section{Step-size convergence}
As indicated in the main manuscript, we use a step size of $\Delta \beta = 0.05$ or smaller in the fourth order Runge-Kutta method while integrating the system of ODE's in the projected CI theory.
In Fig.~\ref{figSI3},  we plot the error in internal energy for the six-site Hubbard models with $U/t = 2$ at half-filling (left panel), and with $U/t=10$ with four electrons (right panel).
The results numerically demonstrate the convergence of the integration with respect to the step size.

\bibliography{CanonTFD}